\documentclass{aa}
\usepackage{graphics}
\begin{document}
\thesaurus{ 
	3(08.05.1; 
	09.08.1; 
	09.09.1 N11C; 
	11.13.1)  
	} 

\def\frac{$''$\hspace*{-.1cm}}
\def\deg{$^{\circ}$\hspace*{-.1cm}}
\def\min{$'$\hspace*{-.1cm}}
\def\h2{H\,{\sc ii}}
\def\hi{H\,{\sc i}}
\def\hb{H$\beta$}
\def\ha{H$\alpha$}
\def\hd{H$\delta$}
\def\heii{He\,{\sc ii}}
\def\hg{H$\gamma$}
\def\sii{[S\,{\sc ii}]}
\def\siii{[S\,{\sc iii}]}
\def\oiii{[O\,{\sc iii}]}
\def\oii{[O\,{\sc ii}]}
\def\hei{He\,{\sc i}}
\def\sm{$M_{\odot}$}
\def\sl{$L_{\odot}$}
\def\ab{$\sim$}
\def\x{$\times$}
\def\sec{s$^{-1}$}
\def\cm2{cm$^{-2}$}
\def\mcube{$^{-3}$}
\def\lam{$\lambda$}

\def\fc{Fe\,{\sc{iii}}}
\def\fb{Fe\,{\sc{ii}}}
\def\hea{He\,{\sc{i}}}
\def\heb{He\,{{\sc ii}}}
\def\nb{N\,{\sc{ii}}}
\def\nc{N\,{\sc{iii}}}
\def\nd{N\,{\sc{iv}}}
\def\ne{N\,{\sc{v}}}
\def\ca{Ca\,{\sc{ii}}}
\def\sib{Si\,{\sc{ii}}}
\def\sic{Si\,{\sc{iii}}}
\def\sid{Si\,{\sc{iv}}}
\def\cb{C\,{\sc{ii}}}
\def\cc{C\,{\sc{iii}}}
\def\cd{C\,{\sc{iv}}}
\def\fc{Fe\,{\sc{iii}}}
\def\p{P\,Cyg}
\def\ac{Al~{\sc{iii}}\ }
\def\na{Na~{\sc{i}}\ }
\def\nf{Ne~{\sc{i}}\ }
\def\ca{Ca~{\sc{ii}}\ }
\def\soc{S~{\sc{iii}}\ }
\def\Sk{Sk--66$^{\circ}$41}

\title{Compact star clusters of the LMC \h2\, region N11\,C\thanks
   {Based on observations obtained at the European Southern 
   Observatory, La Silla, Chile}
}

\offprints{M. Heydari-Malayeri, Mohammad.Heydari-Malayeri@obspm.fr}

\date{Received 4 April 2000 / Accepted 24 July 2000}

\titlerunning{LMC N\,11C}
\authorrunning{Heydari-Malayeri et al.}

\author{Mohammad Heydari-Malayeri\inst{1} \and  
Pierre Royer\thanks{Research Fellow FNRS (Belgium)}\inst{2}  \and 
Gregor Rauw\thanks{Postdoctoral Researcher FNRS (Belgium)}\inst{2} \and 
Nolan R.\ Walborn\inst{3}
}

\institute{{\sc demirm}, Observatoire de Paris, 61 Avenue de l'Observatoire, 
F-75014 Paris, France \and
Institut d'Astrophysique et de G\'eophysique, Universit\'e de Li\`ege, 5, 
Avenue de Cointe, 
B-4000 Li\`ege, Belgium \and
Space Telescope Science Institute, 3700 San Martin Drive, Baltimore, 
MD\,21218, U.S.A.
} 

\maketitle

\begin{abstract} 
Based on imaging and spectroscopy obtained at the ESO NTT telescope
and using an efficient image analysis algorithm, we study the core of
the LMC OB association LH\,13, particularly the two compact stellar
clusters \Sk\, and HNT in the \h2\, region N\,11C.  We resolve \Sk\,
into 15 components and for the first time the HNT cluster into 70
stars, and derive photometry for the members. Moreover, from medium
resolution spectroscopy we determine the spectral types for sixteen
stars in N\,11C.  We compare the color-magnitude diagrams of the
clusters with that of the field stars and discuss the cluster ages.
With an age of \ab\,100\,Myr, the HNT cluster appears
   significantly older than the very young ($\leq 5$\,Myr) \Sk\, 
   starburst. We suggest that most of the `field' O-stars in the 
   core of N\,11C have actually been ejected from \Sk\, through 
   dynamical interactions in the compact cluster. The properties
   of the \Sk\, and HNT clusters suggest that we are viewing different
   star formation regions lying at different distances along the 
   same line of sight.


\end{abstract}

\section{Introduction}

\begin{figure*}[htb]
\resizebox{12cm}{12cm}{\includegraphics{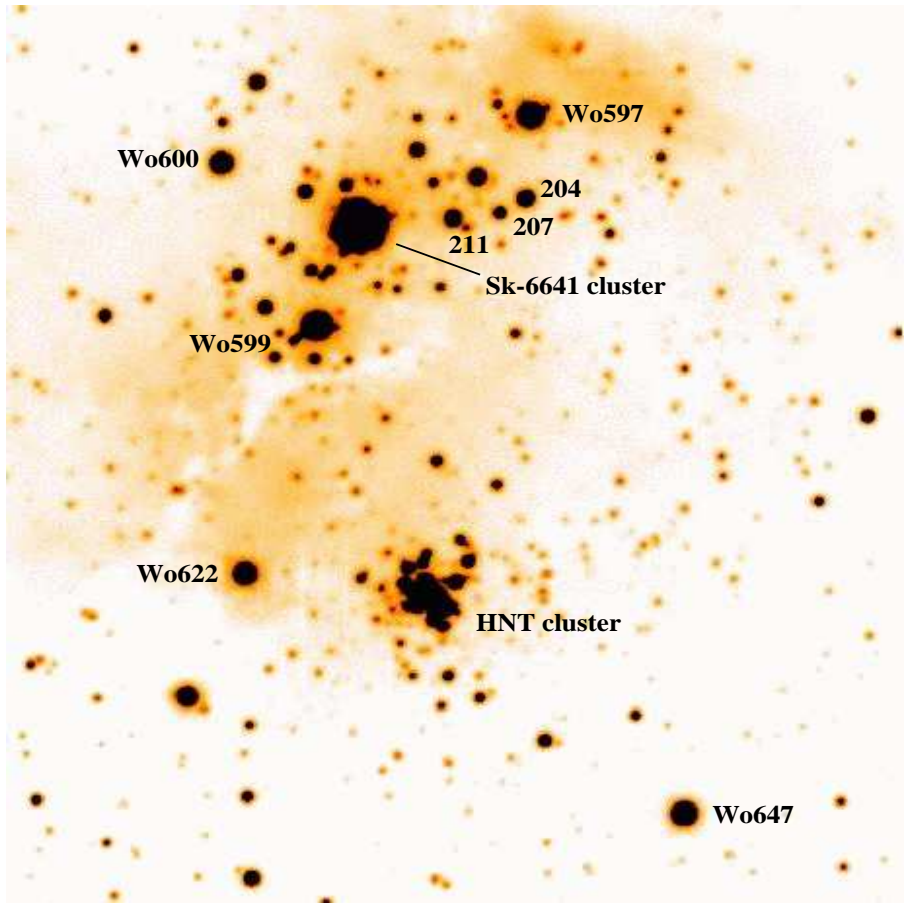}}
\caption{The LMC \h2\, region N\,11C as seen through the Str\"omgren $b$ 
filter. The image is based on co-adding 7 dithered individual exposures. 
The main objects are labelled. The field size is 116\frac\,\x\,127\frac, 
corresponding to \ab\,29\,\x\,32 pc. North is up and east is left.}
\label{n11c}
\end{figure*}

The giant \h2\, region N\,11 (Henize \cite{henize}) or DEM\,34 (Davies
et al.\ \cite{dem}) is the second most important \ha\, emission
complex in the Large Magellanic Cloud (LMC) after the famous 30 Dor
(see e.g.\ Rosado et al.\ \cite{rosado} and references
therein). Interestingly, this region has been suggested to be
reminiscent of an evolved, some 2\,\x\,10$^{6}$ years older version of
the 30 Dor starburst (Walborn \& Parker \cite{wp}). The \h2\, region
N11\,C (NGC 1769), lying at the eastern periphery of the bubble
created by the central association LH\,9 (Walborn \& Parker
\cite{wp}), is one of the brightest and youngest nebular components of
the N\,11 complex.  N11\,C was studied by Heydari-Malayeri et al.\
(\cite{hmnt}, hereafter Paper I) regarding its physical properties
(gas density, excitation, chemical abundances, extinction, etc.) as
well as its stellar content.  They determined the spectral types for 9
stars and gave  $B$ and $V$ photometry for 57 stars in the region.
These observations also revealed the presence of an anonymous compact
cluster (hereafter labelled HNT) south-west of N\,11C
(Fig.\,\ref{n11c}).  Moreover, they identified Wo\,599, an O3--O4
V star, as the main ionizing source of N11\,C, instead of the central
object \Sk\, (HDE\,268743) considered previously to be one of the most
massive stars with a mass well over 120\,\sm\, (Humphreys
\cite{humphreys}). \\

Subsequently, on the basis of sharp images obtained in good seeing
conditions and using high-resolution CCDs assisted by advanced image
restoration methods, Heydari-Malayeri et al. (\cite{hmmr}) showed that
\Sk\, is actually a star cluster made up of at least six components,
the main star having a ZAMS mass of \ab\,90\sm. More recently,
adaptive optics observations at the ESO 3.6\,m telescope resolved
\Sk\, into a tight cluster of at least 12 components, the 
brightest component corresponding to a ZAMS mass of \ab\,50\,\sm\,
(Heydari-Malayeri \& Beuzit \cite{hmb}, hereafter Paper II). \\

The present paper is devoted to the stellar content of N11\,C, which
constitutes the central part of the OB association LH\,13 (Lucke \&
Hodge \cite{lh}). We particularly focus on the two tight star clusters
\Sk\, and HNT.  Although adaptive optics observations have resolved
\Sk\, (Paper II), we need colors, especially in the visible, for
studying the properties of the individual components. With regard to
HNT, no studies have previously been devoted to this cluster.\\

\begin{figure}
\resizebox{\hsize}{!}{\includegraphics{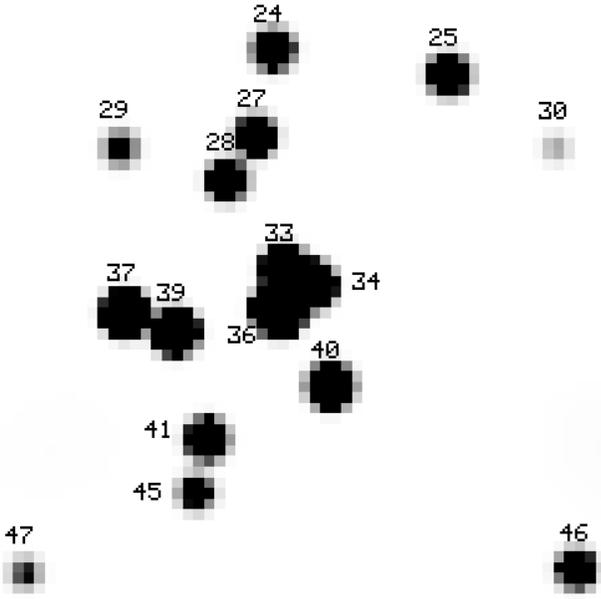}}
\caption{Deconvolved image of the \Sk\, cluster, corresponding to the whole 
blotch appearing in Fig.\,\ref{n11c}. North is up and east is left.
Field 3\frac.9 \x\,3\frac.9. }
\label{sk_core}
\end{figure}

\section{Observations and data reduction}
\subsection{Imaging and  deconvolution}

N11C was observed on 20 November 1997 using the ESO New Technology
Telescope (NTT) equipped with the active optics SUperb Seeing Imager 
(SUSI). The detector was a Tektronix CCD (\#42) with $1024\times1024$
pixels of 24 $\mu$m (0\frac.13 on the sky), and the seeing varied
between 0\frac.56 and 1\frac.14\, ({\sc fwhm}). \\

The observations were performed in the $uvby$ Str\"omgren
photometric system using the ESO filters \# 715, 716, 713, and 714
respectively.  We were particularly careful to keep most of the
brightest stars in the field under the detector's saturation level to
have at our disposal high quality Point Spread Function (PSF)
stars. This led us to adopt exposure times of 180, 130, 150 and 90
seconds in $u$, $v$, $b$ and $y$ respectively. We also used ditherings
of 5\frac\,--10\frac\, for bad pixel rejection and in order to
be able to use the full oversampling capabilities of the MCS
deconvolution algorithm.  Indeed when
performing simultaneous deconvolution of several frames, the algorithm
uses the different frame centerings as a constraint while decreasing 
the pixel size (Magain et al. 1998).
We took a grid of 7 dithering positions for each filter.
Luckily, the objects of interest within N11\,C are close enough to
hold in a single SUSI field of view. Unfortunately, the $u$ images
could not be used for the photometry due to their insufficient
quality.\\

Photometry was derived in the Str\"omgren $v$, $b$ and $y$ filters
according to the following procedure: after bias subtraction and
flat-fielding, the seven frames were co-added in each of the filters.  
The photometry of the stars situated outside of the compact clusters 
was performed on the resulting frames through the PSF fitting
algorithms {\sc allstar} and {\sc daophot} (Stetson \cite{stetson})
implemented in the ESO MIDAS reduction package.  Multiple object
subtraction was performed to clean the images, but only those objects
found during the first activation of the {\sc find} subroutine were
retained for subsequent photometry. This yielded the photometry of 
344 stars lying outside the subfields of the compact clusters \Sk\, and HNT.\\

These clusters were obviously too crowded for {\sc daophot} to work
properly. They were instead processed with the MCS deconvolution algorithm
proposed and implemented by Magain et al. (\cite{magain}).  The
deconvolution was performed on 128\,\x\,128 pixel 
(16\frac.64\,\x\,16\frac.64) sub-frames of the
same co-added frames. Nevertheless, one of the original frames had to
be removed from the sum in the $y$ filter for cluster HNT because of a
badly placed cosmic ray impact, so that only 6 frames were co-added
for that filter before deconvolution of the HNT cluster.  The original
pixel size was reduced by a factor of two for the \Sk\,
deconvolution but it was conserved for the restoration of the
slightly less crowded HNT cluster, thus leading to final PSFs of 0\frac.13 
and 0\frac.26 (FWHM) for the \Sk\, and HNT clusters respectively. \\

The MCS code results from a new approach to
deconvolution taking care not to violate the Shannon (\cite{shannon}) 
sampling theorem: the images are deliberately not deconvolved with the
observed PSF, but with a narrower function, chosen so that the final
deconvolved image can be properly sampled, whatever sampling step is
adopted to represent the final data.  For this purpose, one chooses the
final, well-sampled PSF of the deconvolved image and computes the PSF
which should be used to perform the deconvolution.  The observed PSF
is constructed from several stars close enough to the clusters in
order to avoid any possible PSF variation across the field. \\

\begin{figure*}
\resizebox{9cm}{9cm}{\includegraphics{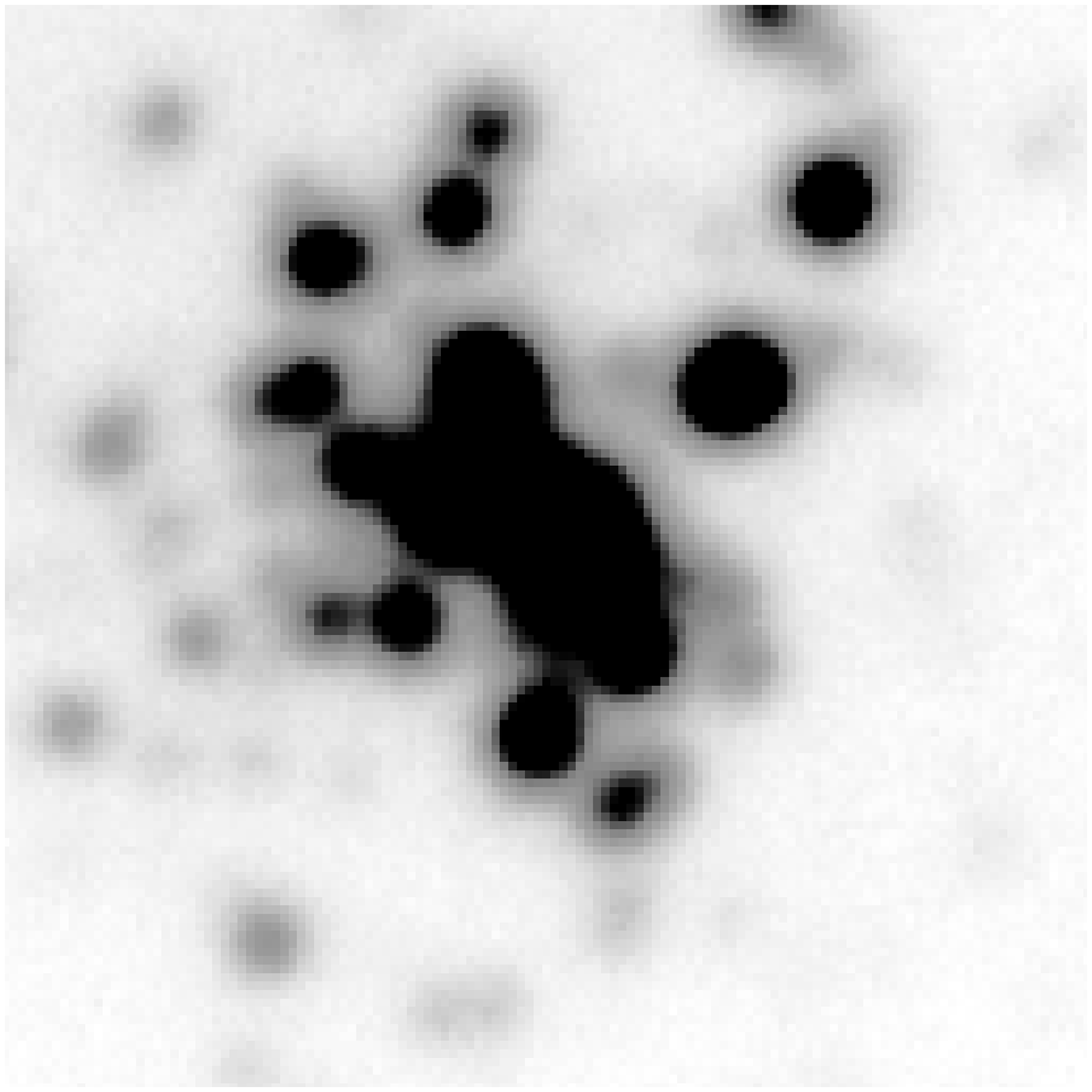}}
\resizebox{9cm}{9cm}{\includegraphics{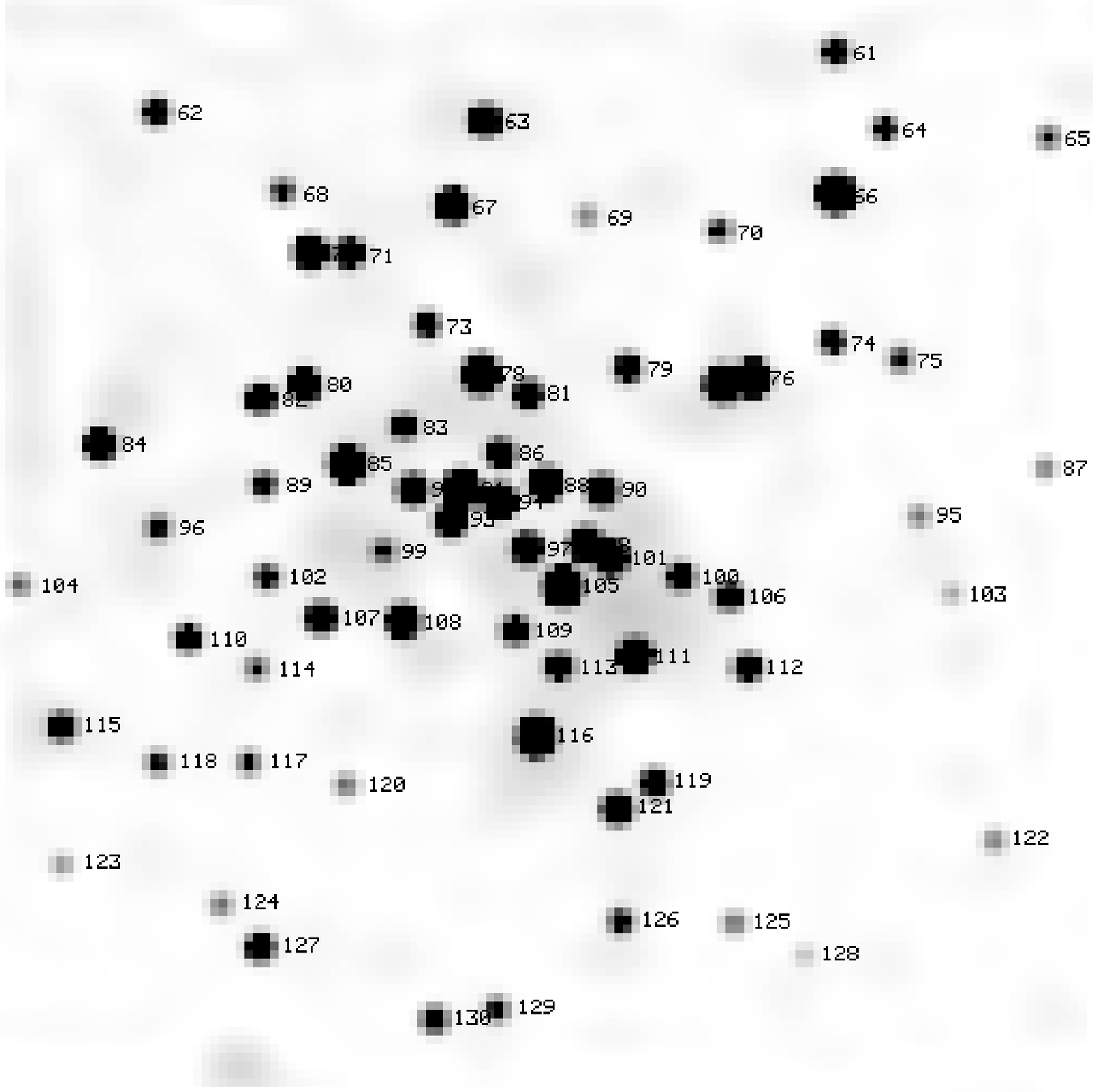}}
\caption{A close-up of the Str\"omgren $b$ image (Fig.\,\ref{n11c}) 
focused on the HNT cluster.
{\bf a}) Original image.
         {\bf b}) Restored image using the MCS deconvolution method.
Field  16\frac.64\,\x\,16\frac.64.}
\label{hnt}
\end{figure*}

The deconvolved frames unveil 63 and 70 objects in and around \Sk\,
and in HNT respectively. Three of the stars in the \Sk\, subfield did
not appear in all filters and were not included in the subsequent
photometric treatment.  Moreover, seven stars in HNT (\#83, 86, 97,
99,103, 113, 128) and three stars in \Sk\ (\#18, 38, 54) were excluded
from further treatment because of suspicious photometry.  The omitted
stars in both cases are exclusively faint components, most of which
appear in or close to the densest and brightest parts of the clusters,
which strongly decreases their already low intrinsic S/N. The final
sample is thus made of 57 stars in the
\Sk\, subfield and 63 in HNT.\\

A technical problem prevented us from using the standard star
observations to calibrate the photometry. Instead we first
deconvolved three bright and isolated stars in the field independently
in order to fix the zero point between deconvolved and {\sc daophot}
photometry. Then, we
fixed the $y$ magnitudes using the $V$ magnitudes published in
Paper~I: out of the 22 stars
of the region around \Sk\, for which these authors published
photometry, 14 fall into our field. The zero point in $y$ was
established by matching our instrumental $y$ magnitudes to the
published $V$ magnitudes for 13 of them, since the fourteenth revealed
a strong discrepancy with respect to the others. The result has an rms
uncertainty of $\pm~0.12$ mag. Finally, we calibrated the $v$ and $b$
magnitudes by making use of the three stars for which we possess both
photometry and spectral type (Wo597, Wo622 and star \#204), by
matching their ($v-y$) and ($b-y$) colors with those calibrated by
Balona (\cite {balona}) for stars with equivalent spectral types. 
The standard deviations of this operation are 0.03 in $v$ and
0.01 in $b$. \\

\begin{figure}
\resizebox{7.8cm}{7.5cm}{\includegraphics{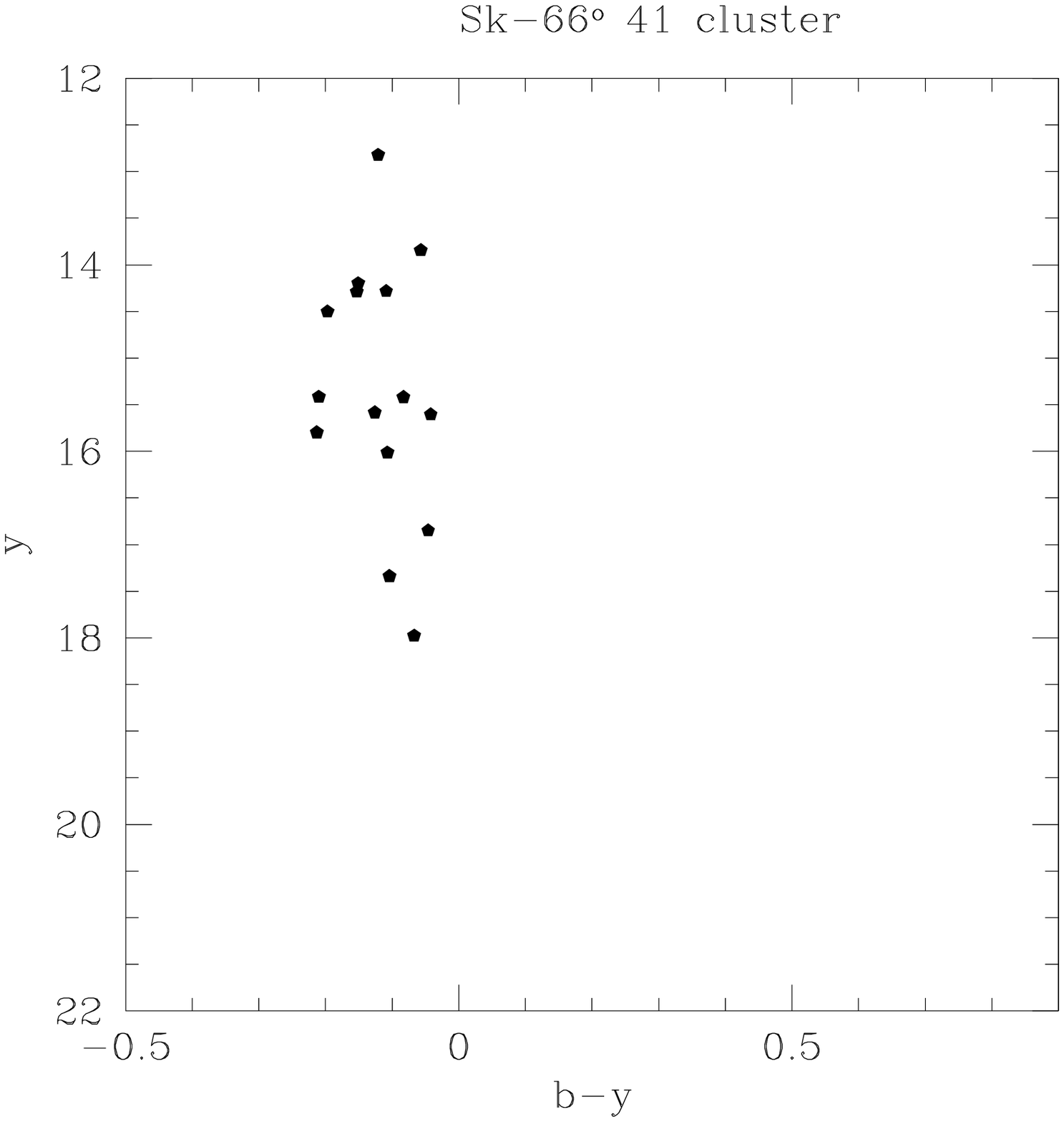}}
\resizebox{7.8cm}{7.5cm}{\includegraphics{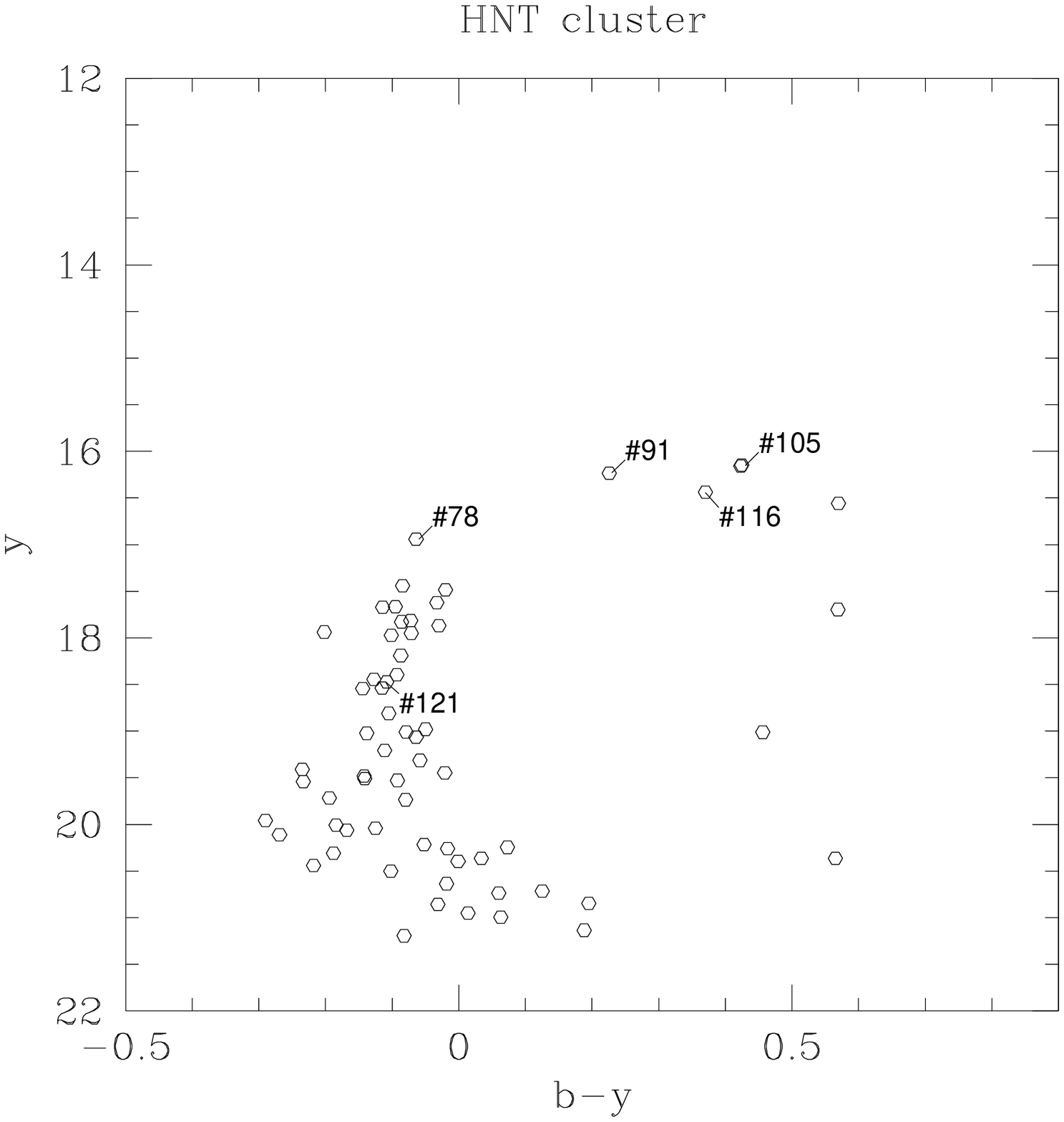}}
\resizebox{7.8cm}{7.5cm}{\includegraphics{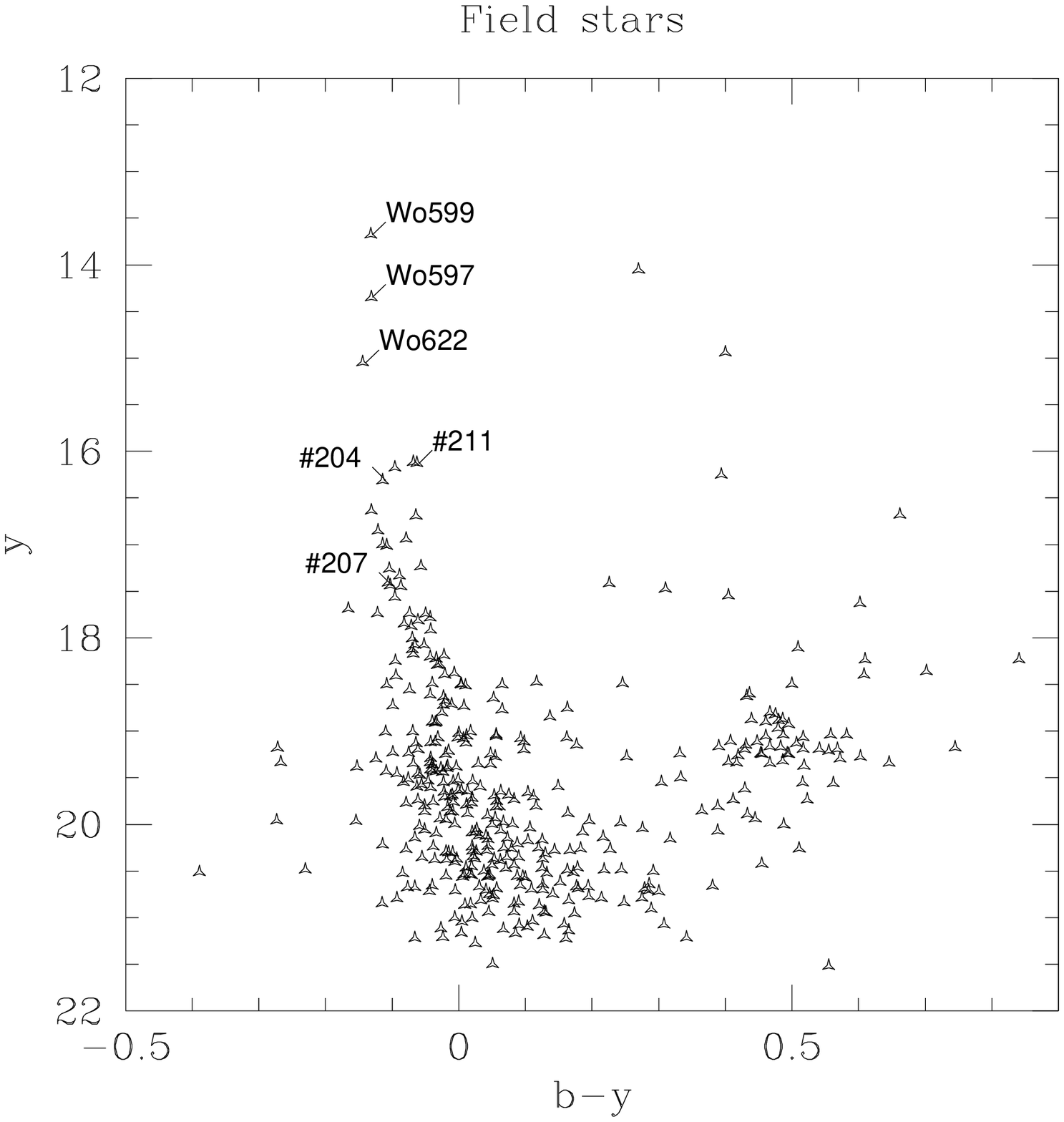}}
\caption{Color-magnitude diagrams of the clusters, \Sk\, and HNT, and 
the field stars. }
\label{cm}
\end{figure}

The final photometric results for the two clusters are presented in
Tables 1 and 2, while the resulting color-magnitude diagrams for the
individual clusters and the field stars are shown in
Fig.\,\ref{cm}. \\

\begin{table}
\caption[]{Photometry of the \Sk\, components. To allow reference to our 
earlier work, the last column provides a cross-identification with the
numbering adopted in Paper II.}
\begin{tabular}{cccccc}   
\hline
Star & $v$   & $b$   & $y$   & $b-y$  & Notes \\
\hline
24   & 15.45 & 15.46 & 15.58 & --0.13 &  \#3 \\
25   & 15.31 & 15.34 & 15.42 & --0.08 &  \#2 \\
27   & 15.40 & 15.56 & 15.60 & --0.04 &  \#4 \\
28   & 15.82 & 15.58 & 15.80 & --0.21 &  \#5 \\
29   & 17.57 & 17.24 & 17.34 & --0.10 & \#12 \\
33   & 14.00 & 14.13 & 14.29 & --0.15 &  \#7 \\
34   & 13.85 & 13.79 & 13.84 & --0.06 &  \#7 \\
36   & 12.64 & 12.70 & 12.82 & --0.12 &  \#8 \\
37   & 14.02 & 14.05 & 14.20 & --0.15 &  \#9 \\
39   & 14.18 & 14.17 & 14.28 & --0.11 & \#10 \\
40   & 14.36 & 14.30 & 14.50 & --0.20 &  \#6 \\
41   & 15.09 & 15.21 & 15.42 & --0.21 & \#11 \\
45   & 17.02 & 16.80 & 16.85 & --0.05 & \#11 \\
46   & 15.91 & 15.91 & 16.01 & --0.11 &  \#1 \\
47   & 18.09 & 17.91 & 17.98 & --0.07 &      \\
\hline
\end{tabular}
\end{table}

\subsection{Spectroscopy with NTT/EMMI}

The EMMI spectrograph attached to the ESO NTT telescope was used on 21
November 1997 (BLMD mode) to obtain several long slit spectra.  
The grating was
\#\,12 centered on \lam 4350\,\AA\, and the detector was a Tektronix CCD
(\#\,31) with 1024$^{2}$ pixels of size 24 $\mu$m.  The range was
\lam\lam 3810--4740\,\AA\, and the dispersion  38\,\AA\,mm$^{-1}$, giving
{\sc fwhm} resolutions of $2.70\pm0.10$ pixels or $2.48\pm0.13$\,\AA\,
for a 1\frac.0 slit. At each position we first took a short 5 min
exposure followed by one or two longer 15 min exposures. The
instrument response was derived thanks to observation of the
calibration stars LTT1020, LTT1788, EG21.  \\

The seeing varied from 0\frac.7 to 1\frac.4. These atmospheric conditions
allowed us to obtain relatively un-contaminated spectra of some of the
components of the HNT cluster, but we were unable to resolve the
more compact \Sk\, cluster spectroscopically.

\section{Photometry results}
 
\subsection{Components of the clusters}

Fig.\,\ref{n11c} presents an image of N\,11C on which the clusters
\Sk\, and HNT as well as the main stars are indicated. The outcomes of
the deconvolution processing applied to these clusters are displayed
in Figs. 2 and 3, while Tables 1 and 2 list the results of the
photometry.\\

The comparison of the present results for \Sk\, with those obtained
using an adaptive optics system (Paper II) is interesting. For this
purpose, Fig.\,\ref{sk_core} has the same size as the
Fig.\,\ref{sk_core} in that paper which displays a deconvolved image
in the near infrared $K$ band. The similarity is almost perfect,
although the present observations reveal 15 components in \Sk\,
instead of 12. The cross identification between the components of the
visible and infrared images is listed in Table 1.  The discrepancy in
the number of components is due to the fact that the images represent
two distinct wavelength ranges and also the clean procedure used for
the $K$ image may be partially responsible. Anyhow, the MCS
deconvolution code reveals that the core of \Sk\, is most probably
made up of at least three components (stars \#33, \#34, and
\#36). These are all hot stars, as indicated by their colors. The
brightest star of the cluster is
\#36 with $y=12.82$. \\

The HNT cluster is apparently richer than \Sk\, but is made up
of fainter stars (Table 2, Fig.\,\ref{hnt}). Only six stars
(\#66, \#78, \#85, \#91, \#105, and \#116) appear brighter than
$y$\,=\,17 mag.  HNT is generally composed of blue stars, although
apart from star \#78, all of the six brightest stars have red
colors.\\

\subsection{Field stars}

The two compact clusters form the densest parts of the LH\,13
association (Lucke \& Hodge \cite{lh}).  Photometry was obtained for
344 stars within the field of LH\,13 but lying outside the clusters.
Several of the brightest ones were also observed spectroscopically, as
presented in Sect. 4.\\

Of particular interest is Wo599. We applied the deconvolution
technique to look into its multiplicity, since it has a composite
spectrum (see $\S$\,4.3). This method did not reveal any close
components towards Wo599.  The closest stars to Wo599, as shown by the
images, are two faint stars of $y=19.68$ and $y=19.38$ lying at
2\frac.6 west and 2\frac.9 east of Wo599 respectively.  This result
does not, however, prove that we are dealing with a single star.

\begin{table*}
\caption[]{Photometry of the HNT cluster}
\begin{flushleft}
\begin{minipage}{5.8cm}
\begin{tabular}{lcccr}   
\hline
Nr. & $v$ & $b$    & $y$    & $b-y$ \\
\hline
61  & 19.77 & 19.52 & 19.72 & --0.19  \\
62  & 19.39 & 19.43 & 19.45 & --0.02  \\
63  & 18.35 & 18.30 & 18.39 & --0.09  \\
64  & 19.97 & 19.82 & 20.01 & --0.18  \\
65  & 20.55 & 20.40 & 20.40 & --0.00  \\
66  & 17.29 & 16.58 & 16.16 &   0.42  \\
67  & 17.79 & 17.74 & 17.83 & --0.09  \\
68  & 20.10 & 20.12 & 20.31 & --0.19  \\
69  & 21.22 & 21.11 & 21.19 & --0.08  \\
70  & 20.43 & 20.24 & 20.26 & --0.02  \\
71  & 19.03 & 18.88 & 19.02 & --0.14  \\
72  & 17.88 & 17.84 & 17.87 & --0.03  \\
73  & 19.80 & 19.67 & 19.96 & --0.29  \\
74  & 19.81 & 19.66 & 19.74 & --0.08  \\
75  & 20.59 & 20.40 & 20.50 & --0.10  \\
76  & 17.64 & 17.57 & 17.67 & --0.10  \\
77  & 17.43 & 17.36 & 17.44 & --0.08  \\
78  & 16.92 & 16.88 & 16.94 & --0.06  \\
79  & 19.55 & 19.37 & 19.51 & --0.14  \\
80  & 18.18 & 18.10 & 18.19 & --0.09  \\
81  & 19.16 & 18.93 & 18.98 & --0.05  \\
\hline
\end{tabular}
\end{minipage}
\begin{minipage}{5.8cm}
\begin{tabular}{lcccr}
\hline
Nr. & $v$ & $b$ & $y$ & $b-y$\\
\hline
82  & 18.73 & 18.70 & 18.81 & --0.11  \\
84  & 19.24 & 18.27 & 17.70 &   0.57  \\
85  & 18.03 & 17.13 & 16.56 &   0.57  \\
87  & 21.31 & 20.93 & 20.36 &   0.57  \\
88  & 17.85 & 17.74 & 17.81 & --0.07  \\
89  & 19.96 & 19.84 & 20.11 & --0.27  \\
90  & 18.61 & 18.40 & 18.54 & --0.14  \\
91  & 16.76 & 16.46 & 16.23 &   0.23  \\
92  & 18.47 & 18.32 & 18.44 & --0.13  \\
93  & 18.00 & 17.88 & 17.95 & --0.07  \\
94  & 17.48 & 17.46 & 17.48 & --0.02  \\
95  & 21.41 & 21.06 & 20.99 &   0.06  \\
96  & 20.54 & 20.22 & 20.44 & --0.22  \\ 
98  & 17.53 & 17.59 & 17.62 & --0.03  \\
100 & 19.56 & 19.31 & 19.54 & --0.23  \\
101 & 17.90 & 17.74 & 17.94 & --0.20  \\
102 & 20.01 & 19.90 & 20.06 & --0.17  \\ 
104 & 21.01 & 20.84 & 20.71 &   0.13  \\
105 & 17.24 & 16.57 & 16.15 &   0.43  \\     
106 & 19.46 & 19.34 & 19.48 & --0.14  \\
107 & 18.49 & 18.42 & 18.54 & --0.12  \\
\hline
\end{tabular}
\end{minipage}
\begin{minipage}{5.8cm}
\begin{tabular}{lcccr}
\hline
Nr. & $v$ & $b$ & $y$ & $b-y$\\
\hline
108 & 17.93 & 17.87 & 17.97 & --0.10  \\
109 & 19.52 & 19.18 & 19.41 & --0.24  \\
110 & 19.29 & 19.26 & 19.31 & --0.06  \\
111 & 17.67 & 17.56 & 17.67 & --0.11  \\
112 & 19.26 & 19.10 & 19.21 & --0.11  \\ 
114 & 21.03 & 20.62 & 20.64 & --0.02  \\
115 & 19.56 & 19.44 & 19.53 & --0.09  \\
116 & 17.49 & 16.81 & 16.44 &   0.37  \\
117 & 20.56 & 20.40 & 20.36 &   0.03  \\
118 & 20.53 & 20.32 & 20.25 &   0.07  \\
119 & 19.17 & 19.00 & 19.07 & --0.06  \\
120 & 20.92 & 20.83 & 20.86 & --0.03  \\
121 & 18.42 & 18.36 & 18.47 & --0.11  \\
122 & 21.63 & 21.04 & 20.85 &   0.20  \\
123 & 21.49 & 21.32 & 21.13 &   0.19  \\
124 & 21.37 & 20.96 & 20.95 &   0.01  \\
125 & 20.98 & 20.80 & 20.74 &   0.06  \\
126 & 20.24 & 19.91 & 20.04 & --0.13  \\
127 & 19.02 & 18.93 & 19.01 & --0.08  \\ 
129 & 20.39 & 20.16 & 20.21 & --0.05  \\
130 & 20.21 & 19.47 & 19.01 &   0.46  \\
\hline
\end{tabular}
\end{minipage}
\end{flushleft}
\end{table*}

\section{Spectral types}

\parindent=0cm

\subsection{\Sk\, cluster}

\begin{figure*}
\resizebox{15cm}{13cm}{\includegraphics{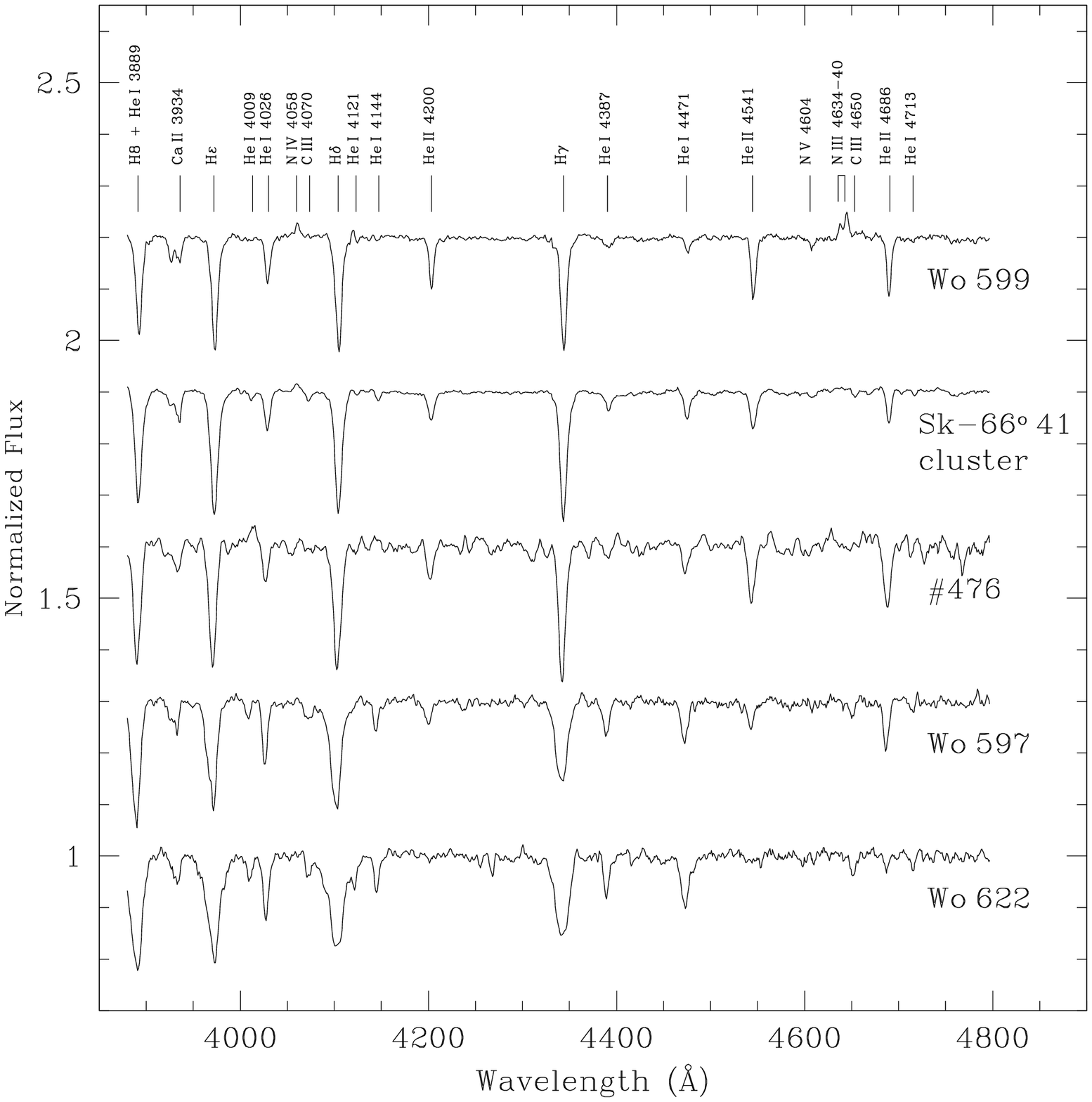}}
\caption{Normalized spectra of the brightest O type stars towards 
N11\,C. That of \Sk\, corresponds to the whole cluster shown 
in Fig.\,\ref{sk_core}. The spectra are shifted
vertically by 0.3 units.     }
\label{m1}
\end{figure*}

It was not possible to obtain the individual spectra of the
components of \Sk.  Therefore the spectrum (Fig.\,\ref{m1}) represents the
whole cluster dominated by the brightest central stars (\#33, \#34,
\#36).  If one were to force a single spectral type, it would be 
O6.5\,V\,((f)).  However, that classification is not satisfactory, 
because there are opposite
discrepancies.  The \nd\,\,\lam 4058 emission (stronger than 
\nc\,\,\lam\lam 4634-40) and \ne\,\,\lam 4604
absorption show the presence of a (non-dominant) O3 component, while
the weaker \hea\, lines such as \lam\lam 4009, 4144, 4387 require a component
at least as late as O8.  Hence, we derive
O3\,V((f*))\,+\,OB.
This result revises the previously published classification
O5\,V, based on lower S/N spectra (Paper I). \\

\subsection{HNT cluster}

\begin{figure*}[htb]
\resizebox{15cm}{13cm}{\includegraphics{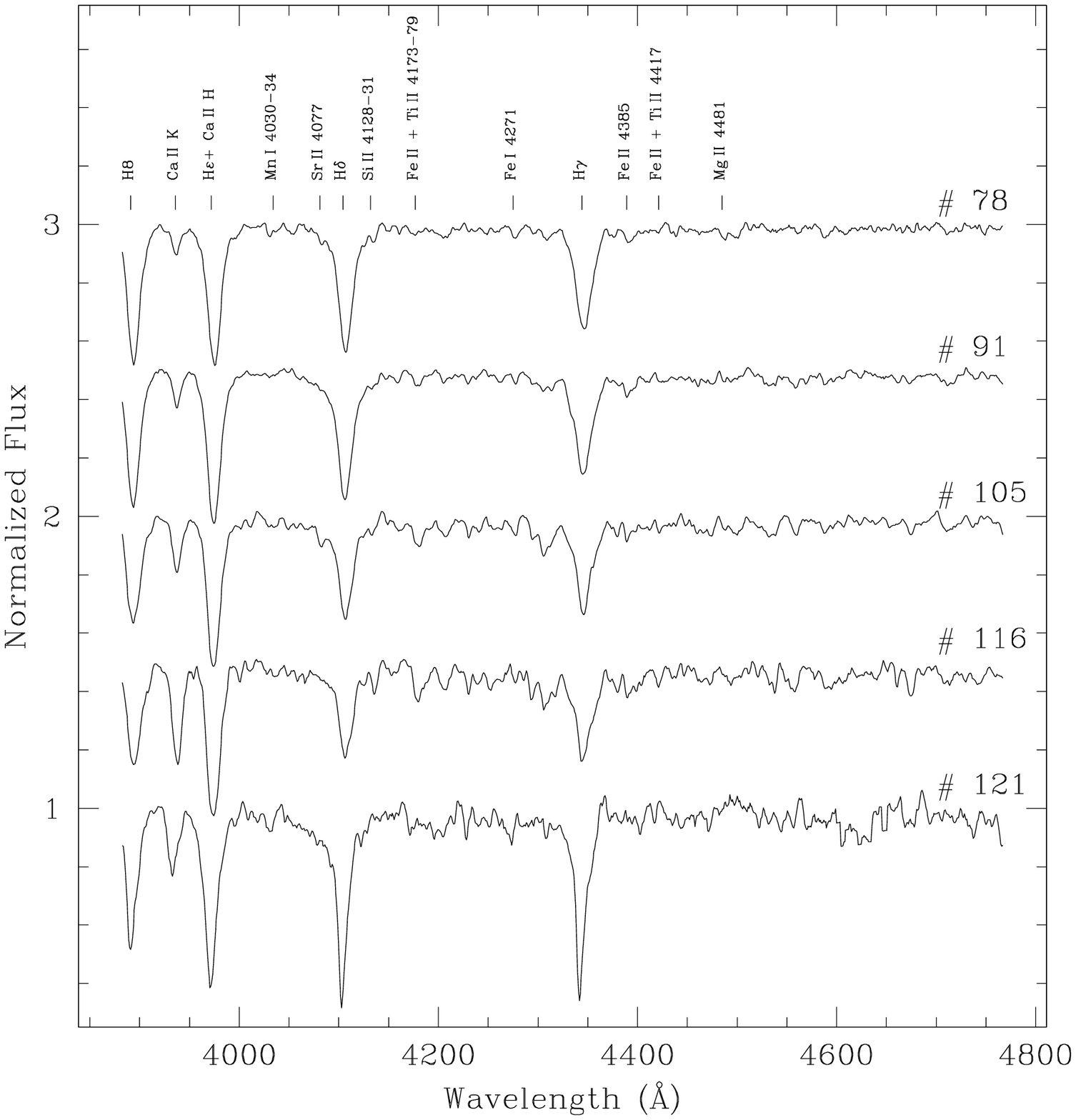}}
\caption{\label{fig: af} Normalized spectra of five of the brightest
members of the HNT cluster. The spectral features that are the most
important for the classification are labelled.}
\label{hnt_spectres}
\end{figure*}

Figure\,\ref{hnt_spectres} displays the spectra of some of the brightest
stars in the HNT cluster. To limit the contamination by nearby fainter
stars, the spectra were extracted over a rather narrow range of four
pixels ($\sim 1.5$\arcsec) centered on the location of the relevant
star along the slit projected on the sky. We caution however that the
resulting spectra are slightly contaminated by the nearby fainter
stars, especially near the crowded cluster center.\\ 

\parindent=1cm

Our spectra of the brighter members of the HNT cluster correspond to
stars of spectral type A-F. We have classified these spectra using the
criteria described by Gray \& Garrison (\cite{GG87}, \cite{GG89a},
b). While these criteria are well established for Galactic stars,
their application to stars in the LMC is more ambiguous because of the
well known metal deficiency of the latter galaxy. Therefore, there is
a dispersion among the classifications based upon different criteria
for the same star. We find that the ratio of the strength of the
\ion{Ca}{ii} K line with respect to the \ion{Ca}{ii} H + H$\epsilon$
line yields an earlier spectral type than the other criteria based for
instance on the strength of the Balmer lines. Since the \ion{Ca}{ii}
lines are weakened compared to Galactic standard stars, we favor the
classification based on the other criteria. Table\,\ref{tab: spect}
lists the results for five bright stars in the HNT cluster. \\

\begin{table*}[htb]
\caption[]{Spectral classification and radial velocities of the stars
in the HNT cluster. The radial velocities are derived from the Balmer
lines excluding the H$\epsilon$ line which is heavily blended with the
\ion{Ca}{ii} H line. The quoted absolute magnitudes are derived from
our photometry assuming an average $E(B-V) = 0.22$ mag. The
last column lists the identification of the fainter nearby stars that
are likely to contribute to the spectra.
\label{tab: spect}}
\begin{tabular}[h]{c c c c c c}
\hline
Star & \multicolumn{2}{c}{Spectral type} & RV & M$_{\rm V}$ & Contam.\\
& \ion{Ca}{ii} & other crit. & km\,s$^{-1}$ & & sources\\
\hline
\#78 & \multicolumn{2}{c}{A0\,--\,A1} & $182 \pm 16$ & $-2.2$ &\#73\\
\#91 & \multicolumn{2}{c}{A0\,--\,A2} & $149 \pm 16$ & $-2.9$ & \#92, 93, 94, 86\\
\#105 & $\sim$ A5 & F0\,--\,F3 & $176 \pm 12$ & $-3.0$ & \#109 \\
\#116 & late A & F3 - F5 & $191 \pm 16$ & $-2.7$ & \\
\#121 & A1 - A2 & A3 - A6 & $160 \pm 32$ & $-0.7$ & \#119\\
\hline
\end{tabular}
\end{table*}

The luminosity classification of A-F stars largely relies on the
intensities of metal lines (e.g.\ Gray \& Garrison \cite{GG87},
\cite{GG89a}, b) and is therefore rather unreliable for the LMC stars
discussed here. We emphasize however that the shape of the wings of
the Balmer lines points towards a giant or even supergiant luminosity
classification. Adopting a distance modulus of 18.5 and assuming an
average $E(B-V) = 0.22$ mag towards N\,11C (Paper I), we derive the absolute
visual magnitudes listed in Table\,\ref{tab: spect}. These values
roughly correspond to luminosity class III\,--\,II. We caution however
that these results might be affected by multiplicity of the stars
considered here.\\ 

We notice that our spectroscopic data reveal no
trace of a star earlier than A0 in the HNT cluster.\\

\parindent=0cm

\subsection{Spectral types of the field stars}

A number of the field stars were studied also spectroscopically.  The
spectrograms of these stars are displayed in Fig.\,\ref{m1}
and Fig.\,\ref{m2}.  Table \ref{tab: ob_velo}
summarizes the spectral types and the measured radial velocities.
Almost all are O or B types. In the following we discuss 
the details of some of the individual spectra. \\

\begin{table*}
\caption[]{Spectral types and radial velocities of the field OB stars}
\label{tab: ob_velo} 
\begin{tabular}{lllllll}   
\hline
Star  	& $y$   & $b-y$ & Velocity   	& Type  & Notes & 
   Alternate ID \\
      	&       &       & km\,s$^{-1}$ 	&       &   &    \\
\hline
\#204 	& 16.31 & --0.11 & 120\,$\pm$\,20 & O8--9:\,V & uncertain & \#2 (Paper I)\\
\#207 	& 17.44 & --0.10 & 124\,$\pm$\,13 & B1--2:\,V & uncertain & \#4 (Paper I)\\
\#211 	& 16.13 & --0.06 & 175\,$\pm$\,38 & B1\,V     &           & \#1 (Paper I)   \\
\#475 	&       &      	 & 160\,$\pm$\,41 & mid B\,I? & uncertain &\\
\#476 	&       &      	 & 124\,$\pm$\,18 & O6\,V     &           &\\  
\#477 	& 17.36 &        & 116\,$\pm$\,66 & B0--1\,V  &           &\\
Wo597 	& 14.35 & --0.13 & 168\,$\pm$\,30 & O9\,V     &           &\\
Wo599 	& 13.68 & --0.14 & 250\,$\pm$\,25 & O4\,V\,((f$^{+}$))\,+\,O & &     \\
Wo622 	& 15.05 & --0.14 & 227\,$\pm$\,29 & O9.7\,III or B0.2\,V      &      \\
\Sk\,  	&       &      	& 326\,$\pm$\,13 &  O3\,V\,((f*))\,+\,OB & cluster & \\
\hline
\end{tabular}
\end{table*}

\parindent=1cm
             
{\underline {{\bf Wo599}}.
This star exhibits an interesting composite spectrum.  Although the
\nd\,\,\lam 4058 emission is prominent, it is weaker than 
\nc\,\,\lam\lam 4634-40.  The presence of \cc\, \lam 4650 
emission is noteworthy. 
The only feature that prevents a pure O4 classification is \hea\,\,\lam 4387,
which is somewhat strange (no other anomalously strong \hea\, lines),
and it is broader than the other lines, which is consistent with an
origin in other star(s).   The required
components are most probably not very late and the earlier type must
dominate, otherwise \hea\,\,\lam 4471 would be stronger. 
We classify Wo599 O4\,V\,((f$^{+}$))\,+\,O. The ``+'' tag means that 
the \sid\,\,\lam\lam 4089, 4116 lines are in emission; when only the latter 
is visible, as in this case, it is presumed that the former is canceled 
by a relatively stronger absorption line. \\

{\underline {{\bf Wo622}}.  The classification of this star is
ambiguous between O9.7\,III and B0.2\,V (both have 
\heb\,\,\lam 4541 $\sim$ \sic\,\,\lam 4552).  The main luminosity criterion
is \heb\,\,\lam 4686/\hea\,\,\lam 4713 and if this is a late O star, it
has to be a giant from that ratio (Walborn \& Fitzpatrick
\cite{wf}). Should the bump at the blue edge of H$\delta$ be \sid\,\,\lam 4089,
the higher luminosity class is supported, but the presence of this
line is uncertain.  On the other hand, if the type is as late as B0.2
then the \heb\,\,\lam 4686/\hea\,\,\lam 4713 ratio would be consistent
with class V.  Given the resolution and S/N, we cannot decide.  Note
however that the derived absolute magnitude (--4.1) is consistent with
the later type (Vacca et al. \cite{vac}).  \\

{\underline {{\bf Star \#211}}. 
The classification is B1\,V, provided that a feature near \heb\,\,\lam
4686 is noise.  If it were real, the type would have to be earlier.\\

{\underline {{\bf Star \#475}}. 
This star lies \ab\,45\frac\, east of Wo622 on a line
joining Wo622 to the middle of HNT, outside the field of our images.
Some features of the spectrum look like those of a mid- or late-B
supergiant, but they are not consistent.  For example, there may be strong 
\cb\,\,\lam 4267, but there is no trace of Mg\,{\sc ii}\,\,\lam 4481, 
which should be very strong.\\

{\underline {{\bf Star \#476}}.
This star lies \ab\,65\frac\, west of \#204  on a line 
joining it to \#211. It is an  O6\,V, assuming that a feature at
\lam 4387 is noise, since there are three other comparable ones 
near H$\gamma$.  Otherwise \hea\,\,\lam 4387 would be too strong, but
everything else is consistent.\\

{\underline {{\bf Star \#477}}.
This star lies \ab\,46\frac\, east of \Sk. Its spectral type is 
B0-1\,V depending on the feature near \heb\,\,\lam 4686.  
If real, the earlier type applies; if not, the later.\\

{\underline {{\bf Wo\,647}}. We classify this star as F7-F8.  
Contrary to what happens for the
members of the HNT cluster, the line intensities in the spectrum of
Wo\,647 match those of the Galactic main sequence or giant standard
stars. This star has a radial velocity of $-127 \pm
11$\,km\,s$^{-1}$. Our results confirm that Wo\,647 is most probably a
Galactic foreground star as already suspected in Paper I.

\begin{figure*}
\resizebox{15cm}{13cm}{\includegraphics{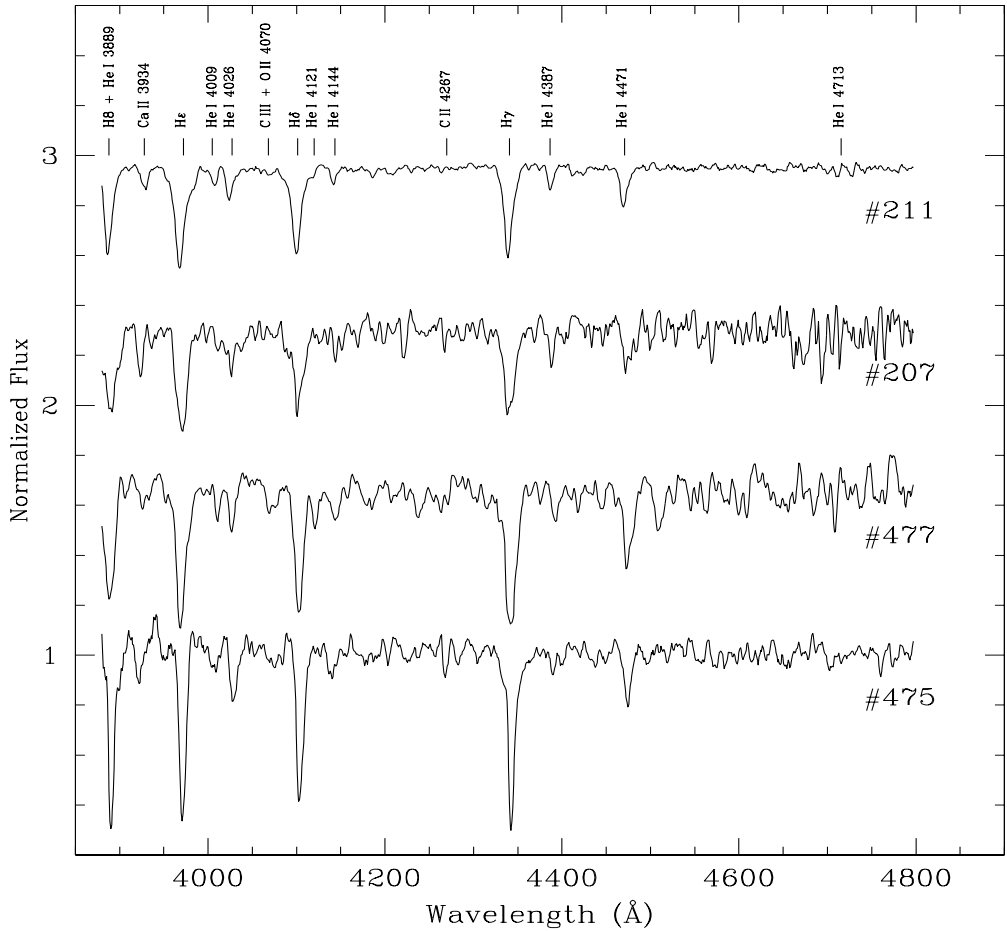}}
\caption{Normalized spectra of B type stars towards N11\,C.}
\label{m2}
\end{figure*}

\begin{figure*}[htb]
\resizebox{13cm}{5cm}{\includegraphics{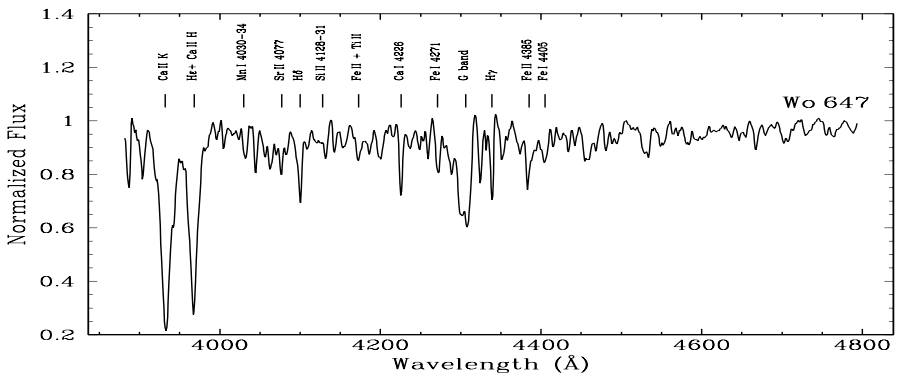}}
\caption{\label{fig: wo647} 
Rectified spectrum of Wo\,647. We classify this star as an F7\,--\,F8
foreground star.}
\label{wo647}
\end{figure*}

\parindent=1cm

\section{Color-magnitude diagrams}

We have derived  color-magnitude diagrams for the two
clusters and the field stars in N11C. Whilst these diagrams are solid
enough for the conclusions reported in this paper, we caution that the
data are not in the standard Str\"omgren system.\\

The C-M diagram of the \Sk\, cluster (Fig.\,\ref{cm}) displays a
rather well defined main sequence in the interval $12.8 \leq y \leq
18.0$ and $-0.2 \leq b - y \leq 0.0$. There is no evidence for stars evolved
off the main sequence in \Sk.\\

Assuming an average reddening of E(B-V) = 0.22 (Paper I) and adopting
a distance modulus of 18.5, we can in principle estimate the age of
the various stellar populations in N\,11C by comparing the observed
C-M diagrams with the isochrones derived from theoretical evolutionary
tracks. However, fitting isochrones to very young stellar populations
such as those of \Sk\, with high turn-off masses and no red
supergiants is a quite difficult and rather uncertain procedure. In
our case, this is even more uncertain since we are dealing with a
non-standard C-M diagram. We have nevertheless attempted a comparison
of our color-magnitude diagrams to the isochrones corresponding to the
Z = 0.008 models of Schaerer et al.\ (\cite{schaerer}). The isochrones
were computed using a program kindly provided by Dr.\ G.\ Meynet. For
the \Sk\, cluster we find that a reasonable upper limit to the age of
the cluster is 5\,Myr. This result is also in line with the spectral
classification of the integrated spectrum of \Sk\,as O3V((f*)) + OB.  
In fact the most massive star of the cluster has not yet left the main
sequence though it shows Of emission features characterizing pre-WR
stages.  Assuming a lowest possible ZAMS mass of 60 \sm\, for an O3
star, this spectral type puts an upper limit on the age of this star
of 3.7 Myr (Schaerer et al. \cite{schaerer}), in agreement with the
upper limit on the cluster age derived from the C-M diagram.  \\

The C-M diagram of the HNT cluster is also shown in Fig.\,\ref{cm}.
The labels indicate those stars for which we have derived spectral
types using our spectroscopic data. The main sequence of the HNT
cluster is visible up to $y \geq 17$. To the right of the main
sequence, we find a couple of evolved stars with $b - y$
between 0.2 and 0.6 that are brighter than $y = 17$.  Applying the
same technique as above to the HNT cluster, we find from the main
sequence turn-off and the red giant population that the age of this
cluster is most probably $100 \pm 25$\,Myr, i.e.\ much older than the
age derived for the \Sk\, cluster.\\

The C-M diagram of the field stars in N\,11C is shown in
Fig.\,\ref{cm}.  Inspection of this diagram reveals a mixture of
stars of quite different ages. We find a main sequence that extends up
to $y \sim 14.3$ as well as a couple of evolved stars to the
right. Around $y = 19$ and $b - y = 0.5$, we notice the clump of older
red giant field stars.\\

The difference in age between the two clusters is an interesting
feature that raises some questions about the physical link between the
HNT cluster and the N\,11C complex. In fact, recent studies have revealed
that massive stars tend to form in a rather coeval fashion. Massey et
al.\ (\cite{Massey}) report results for a sample of 19 OB associations
in the Magellanic Clouds. In about half of these associations, they
find that most of the massive stars formed within a short time ($<
1$\,Myr). In the remaining associations, they found that star
formation most probably occurred over a time-span of less than 10
Myr. Therefore the age difference between \Sk\, and the
HNT cluster appears rather unusual if we assume that both clusters are 
part of the same association. The C-M diagram of the whole set of 464 
stars observed towards LH\,13 indicates that star formation is not coeval. 
This conclusion is in agreement with the results of $UBV$ photometry 
obtained by DeGioa-Eastwood et al.\ (\cite{deg}). A possible explanation 
could be that we are viewing different star formation regions lying at 
different distances along the same line of sight.

\section{Discussion and concluding remarks}

In this work we have obtained sub-arcsecond angular resolutions
(0\frac.13 and 0\frac.26 FWHM) comparable to those of the adaptive
optics owing to the MCS deconvolution algorithm. This technique has
allowed us to push the resolution of the \Sk\, cluster further into 15
components, while at the same time yielding accurate photometry of the
components. It has also enabled us to present a first study of HNT,
the other tight star cluster lying towards the core of the LH\,13
association. With its 70 components, HNT is richer than \Sk, but its
stars are fainter and less massive, the brightest components being
A--F types. \\

\parindent=1cm

It is interesting to note that the {\it ROSAT}-HRI X-ray observations
of Mac Low et al.\,(\cite{mac}) revealed a point-like source in
N\,11C.  Although the number of HRI counts is quite low and Mac Low et
al.\,(\cite{mac}) caution that they cannot confirm the point nature of
this source, its position is in very good agreement with the optical
position of the
\Sk\, cluster. Interestingly the same HRI observations revealed no
X-ray emission associated with the famous tight cluster HD\,32228 at
the core of LH\,9 (south of N\,11B) which contains at least 16
early-type stars with the brightest components being of spectral type
O9\,Ib and O8.5\,II(f) (Walborn et al.\,\cite{wal}, see also Parker 
et al.\,\cite{parker}).  These results suggest that the X-ray emission seen
in N\,11C is most probably due to the interaction of the stellar winds
of the components of the \Sk\, cluster with the relatively dense
ambient interstellar medium, whereas the lack of X-ray emission from
HD\,32228 is due to the lack of a sufficiently dense interstellar
medium in the LH\,9 region. Further constraints on the nature of this
X-ray source will have to await the {\it XMM} observations of the
N\,11 complex. \\

 The \Sk\, cluster harbors a very hot star of spectral type O3 
   and therefore provides the main exciting source of the N\,11C
   \h2\, region contrarily to the finding of Paper I. The ionized 
   gas streaming from N\,11C has a radial velocity of 288 
   km\,s$^{-1}$ (Rosado et al. \cite{rosado}). We have measured 
   mean nebular line radial velocities of 296.0 and 293.6 
   km\,s$^{-1}$ in the spectra around the HNT cluster and Wo599. 
   Given our spectral resolution, these are in very good agreement
   with Rosado et al.'s (\cite{rosado}) results.\\

An inevitable question is whether the two compact clusters, one high
mass the other low mass, belong to the same star formation region.
This question is crucial for better understanding star formation in
the LMC OB association LH13.  The radial velocity derived for the HNT
cluster is 172\,$\pm$\,15 km\,s$^{-1}$ (rms), based on the five
measurements listed in Table
\ref{tab: spect}. Putting aside star \#91, which has a somewhat 
discrepant velocity, yields 177\,$\pm$\,11 (rms) km\,s$^{-1}$. On the
other hand, the global radial velocity of the \Sk\, cluster is
326\,$\pm$\,13 km\,s$^{-1}$ (Table \ref{tab: ob_velo}).  Although it
cannot be ruled out that the multiplicity and the internal motion of
the stars within the \Sk\, cluster could alter the measured radial
velocities, the velocity difference between the two clusters is
probably due to their non-association.  This is in line with the
drastic age difference between the two clusters (see $\S$\,5).  \\

Apart from \Sk, the brightest stars towards LH\,13 are Wo597, Wo599,
Wo600, Wo622, and Wo647. There are two late types among them, Wo600
and Wo647, which are both Galactic F types ($\S$\,4.3 and Paper I).
The remaining three bright stars are O types most probably associated
with LH\,13 and the \h2\, region N\,11C, although we notice that the
measured radial velocities of these stars are slightly less positive
than the radial velocity of the N\,11C nebular lines
(288\,km\,s$^{-1}$, Rosado et al.\ \cite{rosado}) and that of the
\Sk\, cluster.  We conclude that these three stars belong to the
LH\,13 association as deduced from the C-M diagram. Moreover, the
probability that three early type stars not belonging to LH\,13 lie by
chance towards this OB association should be very low. These stars
have probably formed along with \Sk\, during the same burst. \\

Models studying formation of massive stars predict that these stars
should never form in isolation, and that those found in isolation have been
ejected from dense stellar clusters (Bonnell et al. \cite{bon}).
Wo599 has a projected distance of \ab\,15\frac\, (3.8\,pc) from
\Sk. Let us consider that \Sk\, represents the core of the massive
stars resulting from the same starburst and assume that Wo599 is
escaping from its birthplace. Escape velocities can be larger than
200 km\,s$^{-1}$ (Leonard \& Duncan \cite{leonard}, Kroupa
\cite{kroupa}), and if we arbitrarily take 50 km\,s$^{-1}$, Wo599
needs a travel time of 75\,000 years to reach its present position.
Wo597, would need a comparable timespan for its
journey. The most distant candidate, Wo622, has a projected distance of
\ab\,50\frac\, (12.5\,pc), and requires a longer travel time of
250\,000 years.  These travel-time estimates are of course lower
bounds, since we deal with the image on the sky of a three-dimensional
configuration in space.  Inversely, one can calculate the minimum
velocities at which the stars could have reached their current
locations. Assuming a lifetime of 3\,Myr, one gets a lower velocity of
\ab\,4 km\,s$^{-1}$ for Wo622.  One might wonder whether the observed
star density of the \Sk\, cluster is sufficiently high for the
dynamical ejection mechanism to work. However, Leonard \& Duncan
(\cite{LD88}) have shown that binary-binary collisions required to
produce high velocity escapees occur in low density clusters, even
though simple estimates suggest that such interactions are
unlikely. Furthermore, the ejection of the OB stars that we observe
around \Sk\, must have happened during an earlier evolutionary stage
when the cluster was most probably more compact than today (Portegies
Zwart et al.\,\cite{port99}). \\

In summary, our high resolution images reveal two tight clusters with 
significantly different ages within the core of the LH\,13 association.
The physical connection between these two clusters is presently not
clear. The younger one, \Sk\, (age $\leq 3.5 - 5$\,Myr) is most likely
the core of a compact starburst event. The surrounding OB-stars might 
have been ejected from the \Sk\, cluster, probably as a result of dynamical
interaction. The older cluster, HNT (age \ab\,100\,$\pm$\,25\,Myr), 
contains no stars earlier than spectral types A-F and its kinematical 
properties as well as its color-magnitude diagram suggest that this
cluster has no direct connection to the \Sk\, starburst and could rather be
a line of sight object.

\parindent=0cm

\begin{acknowledgements} The restoration of the images to sub-arcsecond 
resolutions and the corresponding photometry was possible only through
the use of the MCS deconvolution algorithm. We are indebted to
Dr. Pierre Magain (Institut d'Astrophysique et de G\'eophysique de
Li\`ege, Belgium) and Dr. Fr\'ed\'eric Courbin (Pontificia Universidad
Cat\'olica, Chile) for making the algorithm available to us and for
advice. The authors express their thanks to Dr. Georges Meynet  
for providing the models of the Geneva group.
PR and GR are greatly indebted to the Fonds National de la
Recherche Scientifique (Belgium) for multiple supports. They are also
supported in part by contract P4/05 ``P\^ole d'Attraction
Interuniversitaire'' (SSTC-Belgium) and by the PRODEX XMM-OM Project.
\end{acknowledgements}

{}

\end{document}